\documentclass[pre,preprint,superscriptaddress,amsmath,amssymb,floatfix]
{revtex4}
\usepackage{graphics}
\usepackage[dvips]{graphicx}
\begin{document}
\title{Field-theoretic description of ionic crystallization in the
restricted primitive model \footnote{Dedicated to Bob Evans on the occasion of his 60th birthday} }
 \author{ A. Ciach}
\address{Institute of Physical Chemistry,
 Polish Academy of Sciences, 01-224 Warszawa, Poland}
 \author{O. Patsahan}
\affiliation{Institute for Condensed Matter Physics of the National 
Academy of Sciences of Ukraine, 1 Svientsitskii Str., 79011 Lviv, Ukraine}
 \date{\today} 
\begin{abstract}
Effects of charge-density fluctuations on a phase behavior of the
restricted primitive model (RPM) are studied within a field-theoretic
formalism. We focus on a $\lambda$-line of continuous transitions
between charge-ordered and charge-disordered phases that is observed
in several mean-field (MF) theories, but is absent in simulation
results. In our study the RPM is reduced to a $\phi^6$ theory, and
a fluctuation contribution to
a grand thermodynamic potential is obtained by generalizing the
Brazovskii approach.  We find that in a presence of fluctuations the
$\lambda$-line disappears. Instead, a  fluctuation-induced first-order
transition to a charge-ordered phase appears in the same region of a
phase diagram, where the liquid -- ionic-crystal transition is
obtained in simulations. Our results indicate that the
 charge-ordered
phase should be identified with an ionic crystal.
\end{abstract} 
\maketitle 
\section{Introduction}

Molten salts, ionic liquids or electrolytes can be described by the
restricted primitive model (RPM), where impenetrable
hard cores of diameter $\sigma$ carry charges with equal magnitude $e$
\cite{stell:95:0,fisher:94:0}.
 In the continuum-space RPM a separation into uniform ion-dilute and
 ion-dense phases with an associated critical point occurs at low
 densities, a transition to an ionic crystal of the CsCl type
 occurs at intermediate densities, and at high densities the fcc
 crystal is stable~\cite{stillinger:60:0}. The above phase-behavior was
 confirmed by recent simulations~\cite{vega:03:0}. 
 In the last decade
 a very intensive debate was focused on critical properties of the RPM
\cite{stell:95:0,fisher:94:0,levin:02:0,kleemeier:99:0,wagner:02:0,wagner:04:0,schroer:06:0,kostko:04:0,luijten:02:0,kim:03:0,ciach:00:0,patsahan:04:1,ciach:06:1}. 
Phase transitions to crystalline phases drew much less attention
\cite{smit:96:0,barrat:87:0} until very recently
\cite{bresme:00:0,vega:03:0}.

In addition to the phase transitions found in simulations, 
 a line of continuous phase transitions 
($\lambda$-line) was found in theoretical studies
\cite{stillinger:60:0,outhwaite:75:0,stell:95:0,stell:99:0,patsahan:03:0,patsahan:04:0,ciach:00:0,ciach:01:1}, except from the mean-spherical (MSA)
 and related
approximations
\cite{leote:94:0,evans:95:0,stell:95:0}.
 Along the $\lambda$-line a decay length of a charge-density
 correlation function, which exhibits exponentially damped
 oscillations on the length scale $\sim \sigma$, diverges. In some
 theories the $\lambda$-line is separated from a first-order
 transition by a tricritical point
 (tcp)\cite{ciach:00:0,ciach:05:0,dickman:99:0,kobelev:02:0}. In
 Ref.~\cite{stillinger:60:0} this line was just rejected as an
 unphysical solution. Indeed, a location of the $\lambda$-line on a
 phase diagram depends strongly on a regularization of the Coulomb
 potential inside the hard core \cite{ciach:03:1,patsahan:04:0}. 
This fact may
 indicate that the $\lambda$-line is an artefact that results from
 approximations made in different theories
 \cite{caillol:04:0,patsahan:04:1}. On the other hand, in
 Ref.~\cite{outhwaite:75:0} it was conjectured that a  divergent correlation length is a signature
 of a crystallization. No quantitative arguments supporting the above
 conjecture were given, however.  Thus, a role of the $\lambda$-line
  in the approximate theories~\cite{stillinger:60:0,outhwaite:75:0,stell:95:0,stell:99:0,patsahan:03:0,patsahan:04:0,ciach:00:0,ciach:01:1} (all of them of
 a mean-field (MF) type), and its existence in the RPM,
 remained unclear~\cite{stell:95:0,patsahan:04:1}.

Renewed interest in the whole phase diagram of the RPM, especially in
the $\lambda$-line and the tcp, is motivated by recent results obtained
for the
lattice RPM (LRPM), where positions of ions are restricted to sites of
different lattices.  On lattices with different symmetries, and/or
with a lattice constant $a$ corresponding to different values of
$\sigma/a\ge 1$, the ions form different periodic patterns at low
temperatures $T$ and/or at high densities $\rho$. Different patterns
correspond to different charge-ordered phases. Transitions between the
high-temperature, charge-disordered phase and the charge-ordered
phases are either continuous or first-order, depending on  details
of a lattice structure
\cite{panag:99:0,ciach:00:0,ciach:01:0,ciach:01:1,ciach:03:0,ciach:04:0,ciach:04:1,diehl:03:0,diehl:05:0,diehl:06:0}.
 In particular, on a simple cubic lattice (sc) with $\sigma/a=1$,
 only an order-disorder transition to a phase with two oppositely
 charged sublattices occurs; this transition is continuous for
 $\rho>\rho_{tc}$, where $\rho_{tc}$ denotes density at the
 tcp. The
 phase separation into dilute and dense, uniform phases is only
 metastable~\cite{stell:99:0,dickman:99:0,ciach:00:0,kobelev:02:0}. Note
 that no continuous transition is predicted by the MSA for the
 LRPM~\cite{hoye:97:0}, in an obvious disagreament with
 simulations~\cite{panag:99:0,dickman:99:0,diehl:03:0,diehl:05:0} and
 exact theoretical predictions~\cite{hoye:97:0}.
 The two types of the
 charge-ordered -- charge-disordered transition
 are shown in Fig.1.
 According to recent simulations~\cite{vega:03:0}, the transition
 lines between the
 liquid and the CsCl crystal are very similar to
 the thick lines
 shown in Fig.1. Note that in contrast to
 close-packed crystals,
 the transition density shows significant
 dependence on temperature.
The above observations raise a question on a relation between the
$\lambda$-line in the continuum space and the charge-ordered --
charge-disordered transitions on the lattice.
\begin{figure}
\includegraphics[scale =0.45]{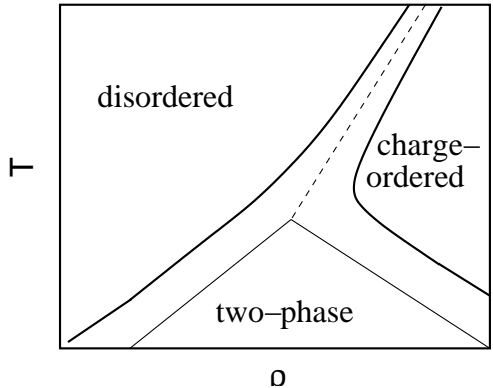}
\caption{Schematic representation of the order-disorder
 transition in the LRPM. Thin dashed- and solid lines represent
 continuous and first-order transitions respectively that were found
 on the sc lattice with $\sigma/a=1$
\cite{panag:99:0,dickman:99:0,ciach:01:1,diehl:03:0,diehl:05:0,ciach:04:0}.
 The dashed line is a
 lattice-analog of the $\lambda$-line. Thick
 solid lines represent the
 first-order transition that occurs for
 $\sigma/a=\sqrt 2,
 2$~\cite{panag:99:0,ciach:04:0,ciach:04:1,ciach:05:0,diehl:06:0}.
 The transition shown by the thin lines can be continuously
 transformed to the transition shown by the thick lines when
 additional nearest neighbor repulsion $J$ is present. For small
 values of $J$ the
 diagram is shown by the thin lines.  When $J$
 exceeds a certain
 value, $J_0$, the dashed line splits into two
 lines that move away
 when $J$ increases
 \cite{ciach:04:0,diehl:05:0}, and the first-order
 transition that
 occurs for large values of $J$ is represented by the
 thick solid
 lines. According to simulation results \cite{vega:03:0},
 the shape
 of the liquid-CsCl crystal two-phase region in
 continuum-space RPM
 is similar to that shown by the solid lines.
 Other transitions that
 occur in some versions of the LRPM and in the
 continuum space at
 low- and at high densities are not shown. $T$ and
 $\rho$ are in
 arbitrary units.}
\end{figure}

In this work we study  effects of fluctuations on the
$\lambda$-line within the field-theoretic description developed in
Ref.\cite{ciach:00:0}. On a MF level of this theory the phase
diagram for each version of the RPM is the same as on the sc lattice
with $\sigma/a=1$ (thin lines in Fig.1). Namely, only the
order-disorder transition that is continuous for $\rho>\rho_{tc}$ is
present
\cite{ciach:00:0,ciach:01:1,ciach:04:0,ciach:04:1}. 
 When fluctuations are
 included  within the
 field-theoretic approch initiated by Brazovskii
\cite{brazovskii:75:0},  in
 some lattice systems the
 order-disorder transition becomes fluctuation-induced first order
\cite{ciach:03:0,ciach:04:0,ciach:04:1} (thick lines in Fig.1). 
The order of the transition agrees 
with simulation results for all considered cases
\cite{panag:99:0,diehl:03:0,diehl:05:0,diehl:06:0}.
 In Ref.\cite{ciach:04:0,ciach:05:0} arguments were given that in the
 continuum-space RPM the order-disorder transition becomes
 fluctuation-induced first order as well. 

Except from the order of the
 considered transition, its location on the phase diagram is of major
 importace for an identification of the charge-ordered phase.  On the MF
 level of our theory the order-disorder transition occurs at low
 densities and high temperatures, and the separation into uniform ion-dilute
 and ion-dense phases is suppresed. 
Beyond MF, and under the
assumption that the order-disorder transition is moved away by
fluctuations, the considered field theory \cite{ciach:00:0} predicts that
for low densities the phase separation into ion-dilute and ion-dense
phases occurs.
 The associated critical point belongs to the Ising
universality-class
\cite{ciach:00:0,ciach:05:0,ciach:06:1}, in agreement with the
earlier theoretical arguments by Stell \cite{stell:92:0,stell:95:0},
and with recent theory \cite{patsahan:04:1}, experiments
\cite{kleemeier:99:0,wiegand:98:0,schroer:06:0} and simulations
\cite{orkoulas:99:0,yan:99:0,luijten:01:0,caillol:02:0,luijten:02:0,kim:03:0}. It is necessary to verify if the
 fluctuations may lead to a shift of the phase boundaries of the
 charge-ordered phase from the phase-space region where the gas-liquid
 separation takes place, to the phase-space region where the CsCl
 crystal is stable, to make the
field-theoretic arguments in favor of the Ising universality class
\cite{ciach:00:0,ciach:05:0,ciach:06:1} complete, and to identify the charge-ordered phase with the CsCl crystal. This is a purpose of our work.

 Our work is based on the 
 Brazovskii theory, which turned out to be
sucessful in a description of
 phase transitions and structure of
soft-matter systems
\cite{fredrickson:87:0,podneks:96:0,levin:92:0}. 
Analogous theory for hard crystals has not been developed yet. The
important common
 feature of the soft- and ionic crystals is that the
periodic ordering
 is not a result of close packing, but follows
directly from
 interaction potentials, or effective, state-dependent
potentials that
 favour periodic structures for any density. Since the leading
physical-mechanism that
 induces the periodic ordering of soft-
 and
ionic crystals is similar, we expect that the Brazovskii approach is
an appropriate description of
 ionic crystallization.

 In sec.2 the field-theoretic description of the RPM is described,
 our
 approximations are discussed, and notation is fixed. In sec.3
 we
 derive approximate expressions for the grand potential with the
 fluctuation-contribution included. The following section is devoted
 to the results obtained for the order-disorder transition. The last
 section contains a short summary and a discussion.

\section{Field-theoretic description of the RPM}
 Field theory for the RPM that is considered in this work was derived in
 Refs.\cite{ciach:00:0,ciach:05:0,ciach:04:1}. In this section we
 summarize the key steps of the derivation, discuss  assumptions
 and approximations, and fix our notation. We consider local
 deviations from the uniform number- and charge densities, $\eta({\bf
 x})=\rho^*({\bf x})-\rho^*_0 =\rho^*_+({\bf x})+\rho^*_-({\bf
 x})-\rho^*_0$ and $\phi({\bf x})=\rho^*_+({\bf x})-\rho^*_-({\bf x})$
 respectively.  $\rho^*_+({\bf x})$ and $\rho^*_-({\bf x})$ correspond
 to a local number-density of  cations and  anions
 respectively, and $\rho^*_0$ is the most probable number density of
 ions. Asterisks indicate that all densities are dimensionless, and
 the unit volume is $\sigma^3$, where $\sigma$ is the core
 diameter. $\phi$ is the charge density in $e/\sigma^3$ units, $e$ is
 the charge.  We focus on systems that are globally charge-neutral,
\begin{equation}
\int_{\bf x}\phi({\bf x})=0,
\end{equation}
where in this paper we use the notation $\int_{\bf x}\equiv\int d{\bf
x}$. Deviations from equilibrium, uniform distributions of ionic
species are thermally excited with the probability density
\cite{ciach:00:0,ciach:06:0,ciach:05:0}
\begin{equation}
\label{pro}
p[\phi,\rho^*]=\Xi^{-1}\exp\big(-\beta \Omega^{MF}[\phi,\rho^*]\big),
\end{equation}
where $\Xi$ is a normalization constant, and in our theory $
\Omega^{MF}$ is approximated by \cite{ciach:00:0,ciach:05:0,ciach:04:1}
\begin{equation}
\label{Omega}
\Omega^{MF}[\phi,\rho^*] =F_h[\phi,\rho^*] + U[\phi] 
-\mu\int_{\bf x} \rho({\bf x}) .
\end{equation}
 $\mu$ is the chemical potential of the ions, $F_h=\int_{\bf x}f_h$ is
 the hard-core reference-system Helmholtz free-energy of the mixture
 in which the core-diameter $\sigma$ of both components is the same. For
 the continuum RPM we adopt the Carnahan-Starling (CS) form of $f_h$ in the
 local-density approximation,
\begin{eqnarray}
\label{fh}
\beta f_h(\rho^*,\phi)=\frac{\rho^*+\phi}{2}
\log\Big(\frac{\rho^*+\phi}{2}\Big)
+\frac{\rho^*-\phi}{2}\log\Bigg(\frac{\rho^*-\phi}{2}\Bigg)-\rho^*
+\rho^{*}\frac{s(4-3s)}{(1-s)^{2}},
\end{eqnarray}
where $\rho^*=\eta+\rho^*_0$, for $\rho^*=\rho^*_0$, the
$\Omega^{MF}[0,\rho^*]$ assumes a minimum, and $s=\pi \rho^*/6$.
Finally, the energy in the RPM is given by
\begin{equation}
\label{URPM}
\beta U[\phi]=\frac{\beta^*}{ 2}
\int_{\bf x}\int_{\bf x'}\theta(|{\bf x'- x}|-1)
 \frac{ \phi({\bf x}) \phi({\bf x}')}{|{\bf x-x'}|}
 =\frac{\beta^*}{2}\int_{\bf k}\tilde \phi({\bf k}) 
 \tilde V (k) 
\tilde \phi(-{\bf k}),
\end{equation}
where $\int_{\bf k}\equiv\int d{\bf k}/(2\pi)^3$. Contributions to the
electrostatic energy coming from overlaping cores are not included in
(\ref{URPM}). We should note that the regularization of the Coulomb
potential for $r<\sigma$ is to some extent arbitrary; in particular, in
Refs.~\cite{patsahan:03:0,patsahan:04:0,caillol:04:0} different
regularizations were
chosen. Here and below  $x=|{\bf x}|$
 is measured in $\sigma$
units. $\beta^*=1/T^*=\beta e^2/(D\sigma)$ is the inverse temperature in
standard reduced units;  $D$ is the
dielectric constant of the solvent. $ \tilde V ( k)=4\pi\cos k /k^2$ is the
Fourier transform of $V(x)=\theta(x-1)/x$, and $k$ is
in $\sigma^{-1}$ units. From the minimum condition for $\Omega^{MF}[\phi,\rho^*]$ we obtain the
relation between $\rho^*_0$ and the intensive parameters,
\begin{equation}
\label{mu}
\beta\mu=\log\rho_{0}^{*}+\frac{s(8-9s+3s^{2})}{(1-s)^{3}}.
\end{equation}

 The fields $\phi$ and $\eta$ occur with the
probability (\ref{pro}), where
 the functional
$\Omega^{MF}[\phi,\rho^*_0+\eta]$ consists of a
 constant term
$\Omega^{MF}[0,\rho^*_0]$ which is irrelevant, and of the term that
depends
 on $\phi$ and $\eta$,
\begin{equation}
\label{delom}
\Delta\Omega^{MF}[\phi,\eta]=\Omega^{MF}[\phi,\rho^*_0+\eta]-
\Omega^{MF}[0,\rho^*_0]\\ 
=\Omega_2[\phi,\eta]+ \Omega_{int}[\phi,\eta].
\end{equation}
The boundary of stability of $\Delta\Omega^{MF}[\phi,\eta]$ 
is determined by the Gaussian part,
\begin{equation}
\label{Gau}
\beta\Omega_2=\frac{1}{ 2}\int\frac{d {\bf k}}{ (2\pi)^3}\Bigg[
\tilde C_{\phi\phi}^0(k)\tilde \phi({\bf k})\tilde\phi(-{\bf k})
+\gamma_{0,2}\tilde\eta({\bf k})
\tilde \eta(-{\bf k})\Bigg]
\end{equation}
where
\begin{equation}
\label{phitad}
\tilde C_{\phi\phi}^0(k)= 
 \rho_0^{*-1}+\beta^* \tilde V(k),
\end{equation}
and
\begin{equation}
\label{Cetad}
 \gamma_{0,2} =
\frac{\partial^2 \beta f_h}{\partial\rho^{*2}}|_{\rho^*=\rho^*_0}=
\frac{1+4s+4s^{2}-4s^{3}+s^{4}}{(1-s)^{4}\rho_{0}^{*}},
\end{equation}
when the CS reference system is used.
The boundary of stability, $\tilde C_{\phi\phi}^0( k_b)=0$,
occurs along the $\lambda$-line 
\begin{equation}
\label{spinodline}
T^*=-\tilde V ( k_b)\rho^*_0\approx 1.61
\rho^*_0,
\end{equation}
 where $ k_b\approx 2.46$ corresponds to the minimum of $ \tilde V( k)$
\cite{ciach:01:1,ciach:04:1,ciach:05:0}. 

The last term in Eq.(\ref{delom}) is local, and can be written as 
\begin{equation}
\label{omloc}
 \beta\Omega_{int}[\phi,\eta]=
\int_{\bf x} \beta\omega_{int}(\phi({\bf x}),\eta({\bf x})),
 \end{equation}
with 
\begin{equation}
\label{omint}
\beta\omega_{int}(\phi,\eta)=\sum_{2m+n>2}\frac{\gamma_{2m,n}}
{(2m)!n!}\phi^{2m}\eta^n,
\end{equation}
where $\gamma_{2m,n}$ are appropriate derivatives of $\beta f_h$.  We
consider a truncated form of $\beta\omega_{int}(\phi,\eta)$, because
otherwise analytical results for the fluctuation-contribution to the
grand potential are not possible. Strictly speaking, the above
expansion can be truncated for $\phi\to 0$ and $\eta\to 0$.  For given
values of $\phi$ and $\eta$, in particular for the results of our
calculations in the ordered phase, however, the truncated expansion
may be oversimplified, especially for small values of $\rho_0^*$ and
for large amplitudes of the fields.

In the field theory
the grand potential and the charge-density correlation function 
are given by 
\begin{equation}
\label{OmegaJ}
\Omega=-kT\log \Xi,
\end{equation} 
and
\begin{equation}
\label{Charge_cor}
\langle \phi({\bf x}) \phi({\bf x}')\rangle
=\Xi^{-1}\int D\eta\int D\phi e^{-\beta\Delta\Omega^{MF}}
\phi({\bf x}) \phi({\bf x}')
\end{equation}
respectively, where
\begin{equation}
\label{Xi}
\Xi=\int D\eta\int D\phi e^{-\beta\Delta\Omega^{MF}}. 
\end{equation}

 In
 the weighted-field approximation (WF) introduced in
Ref.\cite{ciach:00:0}, and described in more detail in
Refs. \cite{ciach:04:1,ciach:05:0,ciach:06:0}, the field
 $\eta({\bf
x})$ is approximated by its most probable form for each
 given field
$\phi({\bf x})$. Another words, for a given field
 $\phi({\bf x})$,
the field $\eta({\bf x})$ is determined by the
 minimum of
$\beta\Delta\Omega^{MF}[\phi,\eta]$ ($\delta
\beta\Delta\Omega^{MF}[\phi,\eta]/\delta\eta=0$), and can be written
in the form 
\begin{equation}
 \eta_{WF}(\phi({\bf x}))=\sum_n\frac{a_n}{n!}\phi({\bf x})^{2n},
\end{equation} 
where the coefficient $a_n$ is given in terms of 
$\gamma_{2m,j}$ such that $m+j\le n+1$~\cite{ciach:04:1}.
 Insertion of $\eta_{WF}(\phi({\bf x}))$ into Eq.(\ref{delom}) 
leads to simplified forms of Eqs.(\ref{Xi}) and (\ref{Charge_cor}),
\begin{equation}
\label{Xieff}
\Xi=\int D\phi e^{-\beta{\cal H}_{eff}[\phi]}
\end{equation}
and
\begin{equation}
\label{Charge_cor_eff}
\langle \phi({\bf x}) \phi({\bf x}')\rangle
=\Xi^{-1}\int D\phi e^{-\beta{\cal H}_{eff}[\phi]}
\phi({\bf x}) \phi({\bf x}')
\end{equation}
respectively, where
\begin{equation}
\label{hef}
\beta{\cal H}_{eff}[\phi]=
\frac{1}{2}\int_{\bf x}\int_{\bf x'}\phi({\bf x})
 C_{\phi\phi}^0({\bf x}-{\bf x}')\phi({\bf x}')
+\sum_{m=2}^{\infty}\frac{{\cal A}_{2m}}{(2m)!}
\int_{\bf x}\phi^{2m}({\bf x}).
\end{equation}
The coefficient ${\cal A}_{2m}$ is given in terms of $\gamma_{2k,n}$
such that $k+n\le m$~\cite{ciach:04:1}.  For the
fluctuation-contribution to the
 average density we obtain
\begin{equation}
\label{etav}
\langle\eta({\bf x})\rangle=\Xi^{-1}\int D\phi 
\eta_{WF}(\phi({\bf x}))e^{-\beta{\cal H}_{eff}[\phi]}=
\sum_n\frac{a_n}{n!}\langle\phi({\bf x})^{2n}\rangle.
\end{equation}
Note that when the expansion in Eq.(\ref{hef}) is truncated at the
term $\propto \phi^{2m}$, then in a consistent approximation the
expansion in Eq.(\ref{etav}) should be truncated at the term $\propto
\langle \phi^{2n}\rangle$ with $n\le m-1$. Otherwise $a_n$ would contain
 the coefficients $\gamma_{2k,j}$ that in $\beta{\cal H}_{eff}[\phi]$ 
are not included.

 We should mention
 that on the sc lattice the WF approximation yields
 quite good results
 for the locations of the continuous
 order-disorder transition, and of the
 tcp \cite{ciach:04:1}. In
 general, for the approximate WF theory we
 cannot expect exact
 dependence of the calculated quantities on
 $\rho^*_0$.

 In this
 work we shall limit ourselves
 to the $\phi^6$ theory, with ${\cal
 H}_{eff}$ approximated by ${\cal
 H}_{WF}$ of the form
\begin{eqnarray}
\label{phi6t}
\beta{\cal H}_{WF}[\phi]=
\frac{1}{2}\int_{\bf x}\int_{\bf x'}\phi({\bf x})
C_{\phi\phi}^0({\bf x}-{\bf x}')\phi({\bf x}')
+\frac{{\cal A}_{4}}{4!}
\int_{\bf x}\phi^{4}({\bf x})+ \frac{{\cal A}_{6}}{6!}
\int_{\bf x}\phi^{6}({\bf x}).
\end{eqnarray}
 The explicit forms of ${\cal A}_4,{\cal A}_6$ in the WF theory are
 given in Appendix A for the CS reference system. The line of
 instability of ${\cal H}_{WF}[\phi]$ ($\lambda$-line) is given in
 Eq.(\ref{spinodline}).  For stability reasons the expansion in
 Eq.(\ref{hef}) can be truncated at the term $\sim\phi^{2n_0}$, if the
 corresponding couplig constant is ${\cal A}_{2n_0}>0$.  In the case
 of the CS reference system we have ${\cal A}_4>0$ for
 $\rho_0^*>\rho_{tc}^*\approx 0.09795$, and ${\cal A}_4<0$ for
 $\rho_0^*<\rho_{tc}^*$. We find ${\cal A}_6>0$ outside the density
 interval $\rho_{tc}^*< \rho_0^* <0.1541$. Negative coupling constants
 were also found for the RPM in Ref.\cite{brilliantov:98:0}. We shall
 not calculate any quantities for $\rho_{tc}^*<\rho_0^*<0.1541$, where
 the functional (\ref{phi6t}) is unstable. Near the above range of
 densities our results are particularly strongly influenced by the
 lack of the terms $O(\phi^8)$, and are less accurate than elsewhere.

 Note that $\tilde C_{\phi\phi}^0( k)$ given in
Eq.(\ref{phitad}) assumes a minimum for $k=k_b>0$, 
and can  be written in the form 
\begin{equation}
\tilde C_{\phi\phi}^0( k)=\beta^*\big(\tau_0+
\Delta \tilde V(k)\big)
\end{equation}
where
\begin{equation}
\label{tau0}
 \beta^*\tau_0=\frac{1}{\rho_0^*}+\beta^*\tilde V(k_b)
\end{equation}
and 
\begin{equation}
\label{delvau}
\Delta \tilde V(k)=\tilde V(k)-\tilde
V(k_b)\simeq_{k\to k_b}v_2(k-k_b)^2+O((k-k_b)^3).
\end{equation}
Near the line of instability 
of ${\cal H}_{WF}$, we have $\beta^*\tau_0\to 0$ (see Eq.(\ref{spinodline})).  Because $\tilde
C_{\phi\phi}^0( k)$  assumes a minimum for
$k=k_b$, the fluctuations $\tilde
\phi({\bf k})$ with $|{\bf k}|\approx k_b$ dominate. If the
fluctuations with $k$ significantly different from $k_b$ are
irrelevant, i.e. for $\tau_0\ll v_2k_b^2$ \cite{brazovskii:75:0}, the
term $O((k-k_b)^3)$ in Eq.(\ref{delvau}) can be neglected, and
\begin{equation}
\label{Cbraz}
\tilde C_{\phi\phi}^0( k)\approx \beta^*\tau_0+\beta^*v_2(k-k_b)^2.
\end{equation}
Eqs.(\ref{Cbraz}) and (\ref{phi6t})  with ${\cal A}_6=0$ are of a
similar form as in the Brazovskii theory \cite{brazovskii:75:0}.  In
the next section we derive an approximate form for the grand potential
in the $\phi^6$-theory (Eq.(\ref{phi6t}) and (\ref{Cbraz})), by
generalizing the Brazovskii approach.

\section{Construction of the Brazovskii-type approximation for the RPM}
 In the charge-ordered phase characterized by a charge-density
profile that is periodic in space, the fluctuating field can be
written in the form
\begin{equation}
\label{fieldsep}
\phi({\bf x})=\Phi({\bf x})+\psi({\bf x}),
\end{equation}
where
\begin{equation}
\Phi({\bf x})=\langle \phi({\bf x})\rangle
\end{equation}
 describes the
ordered phase with a particular symmetry.
In the ordered phase Eq.(\ref{Xi}) can be rewritten in the form
\begin{equation}
\label{Xieff1}
\Xi=\exp\big(-\beta{\cal H}_{eff}[\Phi]\big)
\int D\psi\exp\big(-\beta
{\cal H}_{fluc}[\Phi,\psi]\big),
\end{equation}
where
\begin{equation}
\label{Omsep}
{\cal H}_{fluc}[\Phi,\psi]={\cal H}_{eff}[\Phi+\psi]-{\cal H}_{eff}[\Phi].
\end{equation}
 For the $\phi^6$ theory (Eq.(\ref{phi6t})) we have
\begin{eqnarray}
\label{Omfluc}
\beta{\cal H}_{fluc}[\Phi,\psi]=
\frac{1}{2}\int_{\bf x}\int_{\bf x'}\psi({\bf x})
C_{\phi\phi}^{fluc}({\bf x}-{\bf x}')\psi({\bf x}')    +
\int_{\bf x}C_1({\bf x})\psi({\bf x})+\\ \nonumber
\frac{1}{3!}\int_{\bf x}C_3({\bf x})\psi({\bf x})^3+
\frac{1}{4!}\int_{\bf x}C_4({\bf x})\psi({\bf x})^4+
\frac{1}{5!}\int_{\bf x}C_5({\bf x})\psi({\bf x})^5+
\frac{{\cal A}_6}{6!}\int_{\bf x}\psi({\bf x})^6,
\end{eqnarray}
where explicit expressions for $C_{\phi\phi}^{fluc}$
and $C_i$ are given in  Appendix B.

By inserting Eq.(\ref{Xieff1}) into Eq.(\ref{OmegaJ}) we obtain a
functional of the charge-density distribution $\Phi({\bf x})$,
\begin{equation}
\label{omOm}
-\beta\Omega[\Phi({\bf x})]=-\beta{\cal H}_{eff}[\Phi({\bf x})]+
\log\Big[ \int D\psi\exp\Big(-\beta{\cal H}_{fluc}[\Phi,\psi]\Big)\Big].
\end{equation}
The above equation gives the grand potential in a system with the
charge density constrained to have the form $\Phi({\bf x})$.  For the
theory given by the coarse-grained Hamiltonian (\ref{hef}), or by its
truncated version (\ref{phi6t}), this result is exact. For the charge
distribution given by $\Phi({\bf x})$, the first term in
Eq.(\ref{omOm}) can be directly calculated. In order to obtain an
approximation for the fluctuation contribution, we rewrite ${\cal
H}_{fluc}[\Phi,\psi]$ in the form
\begin{equation}
\label{Har}
{\cal H}_{fluc}[\Phi,\psi]
={\cal H}_{G}[\Phi,\psi]+\Delta{\cal H}[\Phi,\psi]
\end{equation}
where
${\cal H}_{G}[\Phi,\psi]$ is the Gaussian contribution,
\begin{equation}
\label{Haga}
{\cal H}_{G}[\Phi,\psi]=\frac{1}{2}
\int_{\bf x}\int_{\bf x'}\psi({\bf x})C_{\phi\phi}({\bf x}-{\bf x}')
\psi({\bf x}').
\end{equation}
 $C_{\phi\phi}({\bf x}-{\bf x}')$ is inverse to the  exact
charge-density  correlation function, i.e. in   Fourier representation
$\tilde C_{\phi\phi}(k)=1/\tilde G_{\phi\phi}(k)$, 
where
\begin{equation}
\label{def_G}
\tilde G_{\phi\phi}(k)=\frac{\delta^2(\beta\Omega[\Phi])}
{\delta\tilde\Phi(k)\delta\tilde\Phi(-k)}.
\end{equation}
 Next we make an assumption that $\beta\Delta{\cal
 H}[\Phi,\psi]$ can be treated as a small perturbation. When such an
 assumption is valid, we can write
\begin{eqnarray}
\log\Bigg[\int D\psi e^{-\beta({\cal H}_G+\Delta{\cal H})}\Bigg]
= \log\Bigg[\int D\psi 
e^{-\beta{\cal H}_G}\Big(1-\beta\Delta{\cal H}+
O[(\beta\Delta{\cal H})^2]\Big) \Bigg]\\
=\log\int D\psi e^{-\beta{\cal H}_G}+
\log\Big[1-\langle \beta\Delta{\cal H}\rangle_G+
O(\langle \beta\Delta{\cal H}\rangle_G^2)\Big],
\end{eqnarray}
where $\langle ...\rangle_G$ denotes averaging with the Boltzmann
factor $e^{-\beta{\cal H}_G}$. Assuming again that $\langle
\beta\Delta{\cal H}\rangle_G$ is small, we obtain
\begin{eqnarray}
\label{OmHarval}
\beta\Omega[\Phi]=\beta{\cal H}_{eff}[\Phi]-\log\int D\psi
e^{-\beta{\cal H}_G}
+\langle \beta\Delta{\cal H}\rangle_G
+O(\langle \beta\Delta{\cal H}\rangle_G^2).
\end{eqnarray}
 In the uniform phase $\beta\Omega$ is given by the same expression, but
 with $\Phi=0=\beta{\cal H}_{eff}[0]$.
\begin{figure}
\includegraphics[scale=0.5]{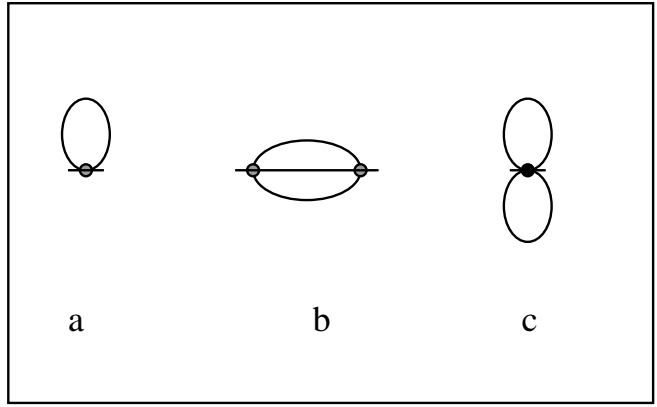}
\caption{Feynman diagrams contributing to $ C_{\phi\phi}$
 in the disordered phase,  to two-loop order.
Shaded circles and a bullet represent ${\cal A}_4$ and ${\cal A}_6$
respectively. Lines represent $ G^0_{\phi\phi}$. In the
self-consistent theory the lines represent $ G^H_{\phi\phi}$. }
\end{figure}

In practice the exact form of $C_{\phi\phi}$ cannot be calculated
analytically. In the perturbation theory
\cite{amit:84:0,zinn-justin:89:0} $ G_{\phi\phi}$ is given by Feynman
diagrams with the $2n$-point vertices ${\cal A}_{2n}$. The vertices at
${\bf x}$ and ${\bf x}'$ are connected by lines representing
$G^0_{\phi\phi}({\bf x}-{\bf x}')$, and all lines are paired. The
corresponding expressions are integrated over all vertex points, or in
Fourier representation over all $ \tilde G^0_{\phi\phi}(k)$-line
loops.  In this work we shall follow the
selfconsistent, one-loop Hartree approximation for $ \tilde
C_{\phi\phi}$ \cite{brazovskii:75:0}.  The one-loop contribution to $
\tilde C_{\phi\phi}$ (Fig.2a) is proportional to
 ${\cal A}_4\int_{\bf k} \tilde
G^0_{\phi\phi}(k)$.  In the effectively
one-loop $\phi^6$ theory (\ref{phi6t}), another contribution to $
\tilde C_{\phi\phi}(k)$ is given by a diagram (Fig.2c) that
 is proportional to
${\cal A}_6(\int_{\bf k} \tilde G^0_{\phi\phi}(k))^2$
\cite{ciach:04:1,ciach:05:0}. The symmetry factors of the graphs are 
calculated according to standared rules \cite{amit:84:0,zinn-justin:89:0}.
 In the selfconsistent, effectively
one-loop approximation $\tilde C_{\phi\phi}(k)$ assumes the approximate form
\cite{brazovskii:75:0,ciach:04:1,ciach:05:0}
\begin{equation}
\label{C_rH}
\tilde C^H_{\phi\phi}(k)=r+\beta^*\Delta\tilde V(k),
\end{equation}
where $r\equiv\tilde C^H_{\phi\phi}(k_b)$, and by using (\ref{def_G}),
(\ref{Omfluc}) and (\ref{Cfluc})-(\ref{c4}) we obtain
\begin{eqnarray}
\label{C_r}
r=
\beta^*\tau_0+
\frac{{\cal A}_4{\cal G}(r)}{2}+
\frac{{\cal A}_6{\cal G}(r)^2}{8}+\frac{1}{2}\Big({\cal A}_4+
\frac{{\cal A}_6{\cal G}(r)}{2}\Big)
\int_{\bf x}\frac{\Phi^2({\bf x})}{V} +
\frac{{\cal A}_6}{4!}\int_{\bf x}\frac{\Phi^4({\bf x})}{V}.
\end{eqnarray}
 In the above  $V=\int_{\bf x}1$ is a volume  of the system and
\begin{equation}
\label{calG0}
 {\cal G}(r)\equiv\langle \psi({\bf x})^2\rangle=
\int_{\bf k} \tilde G^H_{\phi\phi}(k).
\end{equation}

 The remaining diagrams (including the one shown in Fig.2b) are negligible  in the
$\phi^4$ theory for ${\cal A}_4\sqrt{\beta^*v_2}k_b\ll
r$~\cite{brazovskii:75:0}.  When the above condition is not satisified, 
the neglected diagrams, apart from a modification of the form of $r$,  yield
aditional, $k$-dependent
 contribution to $\tilde C^H_{\phi\phi}$ in
Eq.(\ref{C_rH}).  Inclusion
 of such contributions goes beyond the
scope of this work.
 
 In general, the integral in Eq.(\ref{calG0})
cannot be calculated
 analytically. In fact the integral diverges
because of the integrand
 behavior for $k\to\infty$. However, the
contribution from $k\to\infty$
 is unphysical (overlaping hard cores).
When the fluctuations with
 $k\approx k_b$ dominate
($r\ll\beta^*v_2k_b^2$), then the main
 physical contribution to
${\cal G}(r)$ comes from $k\approx k_b$. In
 this case the regularized
integral is \cite{brazovskii:75:0}
\begin{equation}
\label{calG1}
{\cal G}(r)=\int_{\bf k} \frac{1}{r+\beta^*\Delta\tilde V(k)}
\simeq_{r\to 0}\int_{\bf k} \frac{1}{r+\beta^*v_2(k-k_b)^2}=
\frac{2a\sqrt T^*}{\sqrt{r}},
\end{equation}
where 
\begin{equation}
\label{a}
a=k_b^2/(4\pi\sqrt v_2).
\end{equation}
The Eqs.(\ref{C_r}) and (\ref{calG1}) are to be solved
selfconsistently for the ordered and the disordered phases.  In the disordered phase, i.e. for $\Phi=0$,
 $r$ is denoted by $r_0$.

 The second term in  Eq.(\ref{OmHarval}),
 with  $\tilde C_{\phi\phi}(k)$
 approximated  by $\tilde C^H_{\phi\phi}(k)$ (see (\ref{C_rH})), is
\begin{eqnarray}
\label{app}
\log\int D\psi e^{-\beta{\cal H}_{G}}\approx
-2a \sqrt{T^* r}V,
\end{eqnarray}
where the approximation (\ref{delvau}) and the same regularization as in the case of Eq.(\ref{calG1}) were used. For  the last
 term in Eq.(\ref{OmHarval}) we find
(see (\ref{Har}), (\ref{Omfluc}), (\ref{C_rH}) and (\ref{C_r}))
\begin{equation}
\label{flud}
\langle \beta\Delta {\cal H})\rangle_{G}/V=
-\frac{{\cal G}(r)^2}{8}{\cal A}_4-\frac{{\cal G}(r)^3}{24}{\cal A}_6
-\frac{{\cal G}(r)^2}{16}{\cal A}_6\int_{\bf x}\frac{\Phi^2({\bf x})}{V}.
\end{equation}
The above results and Eq.(\ref{calG1}) give the 
 explicit form of the grand potential
 (\ref{OmHarval})
\begin{eqnarray}
\label{diff}
\beta\Omega(\rho_0^*,T^*;\Phi;r]/V=
\beta{\cal H}_{WF}(\rho_0^*,T^*;\Phi]/V+2a\sqrt{rT^* }\\ \nonumber
-\frac{{\cal A}_4a^2}{2}\frac{T^*}{r}
-\frac{{\cal A}_6a^3}{3}\Big(\frac{T^*}{r}\Big)^{3/2}-
\frac{{\cal A}_6a^2}{4}\frac{T^*}{r}\int_{\bf x}\frac{\Phi^2({\bf x})}{V},
\end{eqnarray}
where $r=r(\rho_0^*,T^*;\Phi]$ is a function of $\rho_0^*,T^*$ and a
functional of $\Phi({\bf x})$ that is to be determined from
Eqs. (\ref{C_r}) and (\ref{calG1}).  For given values of $\rho_0^*$
and $T^*$, the value of the (dimensionless) grand potential for a
considered phase corresponds to the minimum of
$\beta\Omega(\rho_0^*,T^*;\Phi;r]$ with respect to $\Phi({\bf x})$,
with $\rho_0^*$ and $T^*$ fixed.  

 $T^*$ represents temperature, but $\rho_0^*$ 
 differs from the average
number-density when the fluctuations are
included (see(\ref{etav})).  The lowest-order fluctuation-induced
density shift, given by the first term in Eq.(\ref{etav}), yields the
leading contribution to the average local density $\rho^*({\bf x})$ of
the form
\begin{equation}
\label{der}
\rho^{*}_1({\bf x})=\rho^*_0-\frac{\gamma_{2,1}}{2\gamma_{0,2}}
\Big({\cal G}(r)+\Phi({\bf x})^2\Big).
\end{equation}
The above gives the density shift in the $\phi^4$ theory.  Higher
order terms in Eq.(\ref{etav}) can be included
 simultaneously with
higher-order terms in Eq.(\ref{phi6t}). In the $\phi^6$ theory the
next-to-leading order term in Eq.(\ref{etav}) leads to the following
approximation for the average density
\begin{equation}
\label{der2}
\rho^{*}_2({\bf x})=\rho^{*}_1({\bf x})+\frac{a_2}{2}\Big[\Phi({\bf x})^4+
6\Phi({\bf x})^2{\cal G}(r) +3{\cal G}(r)^2\Big],
\end{equation}
where $a_2$ is expressed in terms of $\gamma_{2m,n}$ in Appendix A.
The thermodynamic density is given by the space-averaged density
profile according to
\begin{equation}
\label{sp_av}
 \rho^*=\int_{\bf x}\frac{\rho^*({\bf x})}{V}= \int_{V_u}
\frac{\rho^*({\bf x})}{V_u},
\end{equation}
where the integration $\int_{V_u}$ is over the unit cell of the
ordered structure, and $V_u$ is the unit-cell volume. Explicit
expression for the average density in the liquid is given in Appendix
A.

In practice a determination of the equilibrium charge-density profile
$\Phi({\bf x})$ and the phase transition between the charge-ordered
and charge-disordered phases from Eqs.(\ref{diff}), (\ref{C_r})
and (\ref{calG1})
is difficult. The problem simplifies greatly when a form of
$\Phi({\bf x})$ is limited to a particular function that depends on
several parameters.  In this case the functionals are reduced to functions of
several variables, and the problem of obtaining a minimum of
$\beta\Omega$ for a given class of functions becomes tractable.

 For an ordered phase of a particular symmetry, $
\Phi({\bf x})$ can be written as a linear combination of  functions
$g_i({\bf x})$ forming a corresponding orthonormal basis
\cite{podneks:96:0},
\begin{equation}
\label{phi_sum}
\Phi({\bf x})=\sum_i\Phi_ig_i({\bf x}),
\end{equation}
where $\Phi_i$ are the corresponding amplitudes.
In Fourier representation the basis functions can be written in the form
\begin{equation}
\label{g_i}
\tilde g_i({\bf k})=\frac{ (2\pi)^{d}}{\sqrt{2n_i}}
\sum_{j=1}^{n_i}\Big(\delta({\bf k}-{\bf k}^j_{ib})+
\delta({\bf k}+{\bf k}^j_{ib})\Big),
\end{equation}
where for the considered symmetry $2n_i$ and ${\bf k}^j_{ib}$ are the
number of vectors and the $j$-th vector in the i-th shell
respectively. In order to specify the structure we need to know the
vectors ${\bf k}^j_{ib}$ that determine the size of the unit cell of
the structure, apart from the amplitudes $\Phi_i$. We assume that the
vectors forming the first shell correspond to the wave-vectors of the
most probable excitations in the uniform phase.  In the theory
outlined above such wave-vectors are determined by a minimum of
$\tilde C_{\phi\phi}(k)$, since it yields a maximum of the probability
$\propto \exp(-\beta{\cal H}_G)$. In the one-loop approximation
$\tilde C_{\phi\phi}(k)$ assumes a minimum for $|{\bf k}|=k_b$, we
thus assume $|{\bf k}^j_{1b}|=k_b$.

 When the form of $\Phi({\bf x})$ is restricted to Eq.(\ref{phi_sum})
 and $|{\bf k}^j_{1b}|=k_b$, then for given $\rho^*_0$ and $T^*$,
 $r=r(\rho_0^*,T^*;\Phi]$ and $\beta\Omega(\rho_0^*,T^*;\Phi;r]$
 become functions of the amplitudes $\Phi_i$.  Typically, only the
 first one-
\cite{brazovskii:75:0,fredrickson:87:0} or two shells
\cite{podneks:96:0} in Eq.(\ref{phi_sum}) are taken into account in
studies of the fluctuation-induced first-order phase transitions
\cite{brazovskii:75:0,fredrickson:87:0}.  
Our explicit results are obtained in the one-shell approximation,
\begin{eqnarray}
\label{phii}
\Phi({\bf x})=\Phi_1g_1({\bf x}).
\end{eqnarray}
For a few points we also considered the two-shell approximation, but
the long formulas will not be given here.
In the one-shell approximation we can write
\begin{eqnarray}
\int_{\bf x}\frac{\Phi({\bf x})^2}{V}=\Phi_1^2 ,\hskip1cm
\int_{\bf x}\frac{\Phi({\bf x})^4}{V}=
\Phi_1^4s_o,\hskip1cm \int_{\bf x}\frac{\Phi({\bf x})^6}{V}=
\Phi_1^6\kappa_o,
\end{eqnarray}
 where
\begin{eqnarray}
\label{skappa}
s_o=\int_{\bf x}\frac{g_1({\bf x})^4}{V}=\int_{V_u}\frac{g_1({\bf x})^4}{V_u}\hskip1cm\rm{ and}
 \hskip1cm\kappa_o=
\int_{\bf x}\frac{g_1({\bf x})^6}{V}=\int_{V_u}\frac{g_1({\bf x})^6}{V_u}
\end{eqnarray}
 are the geometric factors associated with a particular
symmetry of the ordered phase. 
 In the one-shell approximation the above relations should be inserted
 into Eqs.(\ref{C_r}) and (\ref{diff}), and the extremum
 condition for $\beta\Omega$ can be written in the explicit form
\begin{eqnarray}
\label{min_cond}
r+\Big({\cal A}_4+
a{\cal A}_6\sqrt{\frac{T^*}{r}}\Big)\Big(\frac{s_o}{3!}-
\frac{1}{2}\Big)\Phi_1^2+\frac{{\cal A}_6}{5!}\big(\kappa_o-5s_o\big)
\Phi_1^4=0.
\end{eqnarray}
The resulting set of equations can be solved for each point
$(\rho^*_0,T^*)$ with respect to $r$ and $\Phi_1$. When the results
are inserted in Eq.(\ref{diff}), the minimum of the grand-potential is
obtained for a given pair of $s_o$ and $\kappa_o$, i.e. for a chosen
structure of the ordered phase. The above method of obtaining the
grand potential is equivalent to the method used in
Ref.\cite{brazovskii:75:0,fredrickson:87:0} and outlined in Appendix
D. We verified our calculations by comparing the results obtained by
both methods.



\section{ Transition between the charge-ordered
 and charge-disordered phases}

Let us focus on effects of fluctuations on the $\lambda$-line, which
on the MF level is given in Eq.(\ref{spinodline}). At the $\lambda$-line
the system becomes unstable with respect to the dominant
charge-density fluctuations with the wave number $k_b>0$. At the
boundary of stability, the second functional-derivative of the
grand-potential functional (\ref{OmHarval}) 
is $\tilde C_{\phi\phi}(k_b)=0$. In the
effectively one-loop self-consistent Hartree approximaiton we have
$\tilde C^H_{\phi\phi}(k_b)=r_0$, where $r_0$ is a self-consitent
solution of Eqs.(\ref{C_r}) and (\ref{calG0}) with $\Phi=0$. For $r_0\ll 1$
 the
approximation (\ref{calG1}) is valid, and we easily find that $r_0=0$ is a
solution of Eqs.(\ref{C_r}) and (\ref{calG1}) with $\Phi=0$ only for
$T^*=0$. Thus, in a presence of fluctuations the $\lambda$-line
disappears.

  For a given thermodynamic state there may exist one or several
  minima of $\beta\Omega$ (Eq.(\ref{diff})), associated with ordered
  phases with different symmetries.  The lowest value of the grand
  potential for given $\rho^*_0, T^*$ corresponds to the stable
  phase. Other phases are metastable (unstable) if a minimum of the
  grand potential exists (does not exist).  At the phase coexistence
  two minima of the thermodynamic potential, corresponding to the two
  coexisting phases, are equal; other minima, if present, are
  associated with larger values of the grand potential.

When the expansion for $\Phi({\bf x})$ in Eq.(\ref{phi_sum}) is truncated at the first
shell, analytical results for the phase coexistence are possible.
However, only a very crude approximation for the ordered structure can
be obtained. In this study we shall limit ourselves to analytical
calculations in the one-shell approximation. More accurate results for
the structure of the ordered phase can be obtained numerically in a
future work.
 
 We are interested in a stability of the ionic crystal.
In the case of the CsCl symmetry ($Pn3m$), the first shell is formed
by the three vectors ${\bf k}_b^1/k_b=(1,0,0),{\bf k}_b^2/k_b=(0,1,0)$
and ${\bf k}_b^3/k_b=(0,0,1)$, i.e. $n_1=3$, and in real space
\begin{equation}
\label{gP}
g^{P}_1(x,y,z)=
\frac{2}{\sqrt 6}\Big(\cos(k_bx)+\cos(k_by)+\cos(k_bz)\Big).
\end{equation}
 The above first shell determines the so called P
 structure. Unfortunatelly, topological properties of P differ from
 that of the CsCl crystal. Namely, the P structure is bicontinuous,
 i.e.  the surface $g_1({\bf x})=0$ separates space into the
 positively and negatively charged regions, and both regions are
 continuous, as shown in Fig.3. In the ionic crystal, however, the
 positively and negatively charged regions are topologically eqivalent
 to spheres separated by the uncharged solvent of nonvanishing
 volume. In addition, the nn distance in the P structure is $\sqrt
 3\pi/k_b\approx 2.2$. This distance is much larger than in the actual
 CsCl crystal.   The nn distance in the ionic crystal is closer to the nn distance,
 $\pi/k_b\approx 1.27$,  in the  case of the
  one dimensional ordering (lamellar phase), where $n_1=1$ and
\begin{equation}
\label{glam}
g^{lam}_1(x)=\sqrt 2 \cos(k_bx).
\end{equation}
Since the precise structure of the crystalline phase cannot be
determined within the one- or two-shell approximation, we consider
both phases to find and compare the transition lines between them and
the disordered phase. In this way we gain some insight in the
approximate location of the actual phase transition.

\begin{figure}
\includegraphics[scale=1]{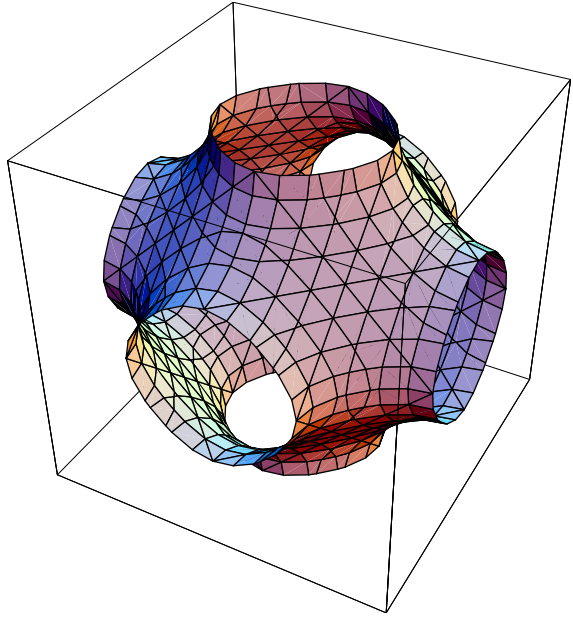}
\caption{The surface $g_1^P({\bf x})=0$ in the unit cell
 of the periodic structure P.  This surface separates the positively
 and negatively charged regions. The lattice constant is
 $2\pi/k_b\approx 2.55$ in ion-diameter units.  }
\end{figure}

\subsection{MF approximation}
In the MF the fluctuation contribution to
(\ref{OmHarval}) is neglected.
  In  the one-shell approximation the grand potential is a function
of the amplitude $\Phi_1$,
\begin{eqnarray}
\label{MFexp}
\beta{\cal H}_{WF}(\rho_0,T^*,\Phi_1)/V=
\frac{1}{2}\beta^*\tau_0\Phi_1^2
+\frac{{\cal A}_4}{4!}s_o\Phi_1^4+\frac{{\cal A}_6}{6!}\kappa_o\Phi_1^6,
\end{eqnarray}
where  the geometric factors $s_o$
and $\kappa_o$ are found to be 
\begin{eqnarray}
s_o=\left\{ 
\begin{array}{ll}
3/2 &\;\; \mbox{lamellar structure}\\
5/2 &\;\; \mbox{P structure },
\end{array}
 \right.
 \label{so}
 \end{eqnarray}
and 
\begin{eqnarray}
\kappa_o=\left\{ 
\begin{array}{ll}
5/2 &\;\; \mbox{lamellar structure}\\
155/18 &\;\; \mbox{P structure }.
\end{array}
 \right.
 \label{kappao}
 \end{eqnarray}
 For
$\rho^*_0>\rho^*_{tc}$ the order-disorder transition is continuous and
coincides with the line of instability (\ref{spinodline}), whereas for
$\rho^*_0<\rho^*_{tc}$ the order-disorder transition is first order,
and occurs when the condition
\begin{eqnarray}
\label{trans_MF}
\frac{\partial \beta{\cal H}_{WF}(\rho_0,T^*,\Phi_1)}{\partial\Phi_1}
=0=\beta{\cal H}_{WF}(\rho_0,T^*,\Phi_1)
\end{eqnarray}
is satisfied.
 The
expressions for the transition lines $T^*_{lam} $ and $T^*_{P}$, and the amplitude $\Phi_1$
are given in Appendix C (Eqs.(\ref{trans_MF_line}) and
(\ref{phi1a})).
It turns out that the P phase is only metastable. However, the
relative difference $ (T^*_{lam}-T^*_P)/ T^*_{lam}$ is very small.

The density in the ordered phase can be obtained from
Eq.(\ref{der})
by  neglecting the fluctuation contribution.
 In the lowest nontrivial order we have
\begin{equation}
\label{rho_space}
\rho^*({\bf x})=\rho^*_0
-\frac{\gamma_{2,1}}{2\gamma_{0,2}}\Phi({\bf x})^2.
\end{equation}
We verified that Eq.(\ref{rho_space}) yields $\rho^*_{\pm}({\bf
x})=(\rho^*({\bf x})\pm\Phi({\bf x}))/2\ge0$ for all space positions.
The space-averaged density in the ordered phases is given in Eq.({\ref{sp_av}).
 The explicit expression for
$\Delta\rho^*=\rho^*-\rho^*_0 $ is given in  Appendix C. The
resulting density-temperature phase diagram is shown in Fig.4.

\begin{figure}
\includegraphics[scale=0.45]{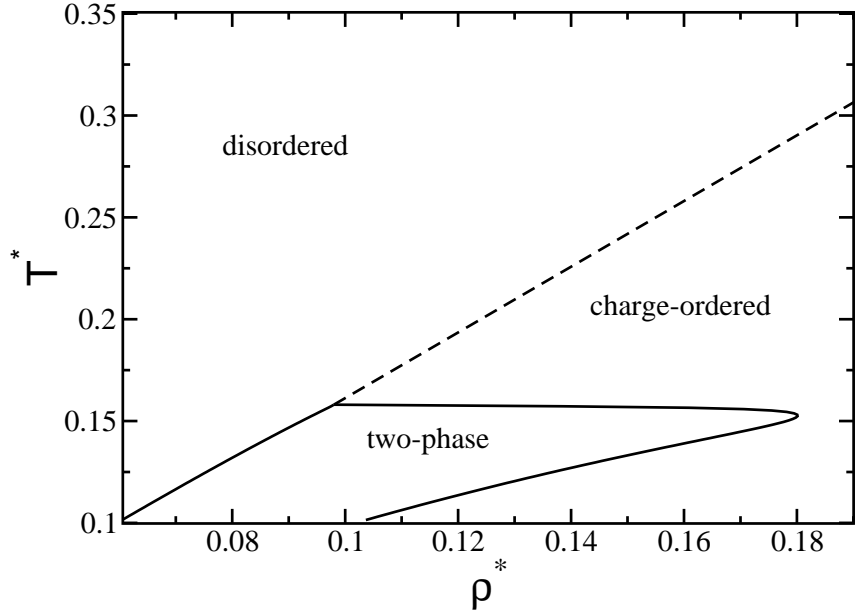}
\caption{Density-temperature MF phase diagram obtained from the
approximate functional (\ref{MFexp}). Temparature $T^*$ and density$\rho^*$ 
are in standard reduced units defined in sec.II.}
\end{figure}
The diagram shown in Fig.4 is obtained with the functional
$\beta\Delta\Omega^{MF}[\phi,\eta]$, Eq.(\ref{delom}), approximated
by
 the functional ${\cal H}_{WF}$, Eq.(\ref{phi6t}). In order to
verify
 the accuracy of the approximation (\ref{phi6t}), we calculated
the
 functional $\beta\Delta\Omega^{MF}[\Phi(x), \rho^*(x)-\rho^*_0]$
along
 the line $T^*_{lam}(\rho^*_0)$, for the fields $\Phi(x)$ and
$\rho^*(x)-\rho^*_0$ that yield ${\cal H}_{WF}=0$. The result shown
in
 Fig.5 indicates that our approximate functional (\ref{phi6t})
yields
 the poorest accuracy in this part of the phase diagram where
${\cal
 A}_6$ is very small (see the discussion below
Eqs.(\ref{omint}) and (\ref{phi6t})), and
 the term $\propto\phi^8$
should be included.  We also considered the
 two-shell approximation
for $\Phi({\bf x})$. We found very similar
 results, with somewhat
lower transition temperatures, and with a
 smaller difference between
them. We conclude that the
 transition-temperature obtained in the
approximate theory (Eq.(\ref{phi6t})) is
 overestimated.
 
 Our main
concern in this work is to determine the
 fluctuation contribution to
the grand potential in
 Eq.(\ref{OmHarval}). We shall
 not attempt to
find better MF results by numerical minimization of the
 functional
(\ref{delom}).
\begin{figure}
\includegraphics[scale=0.45]{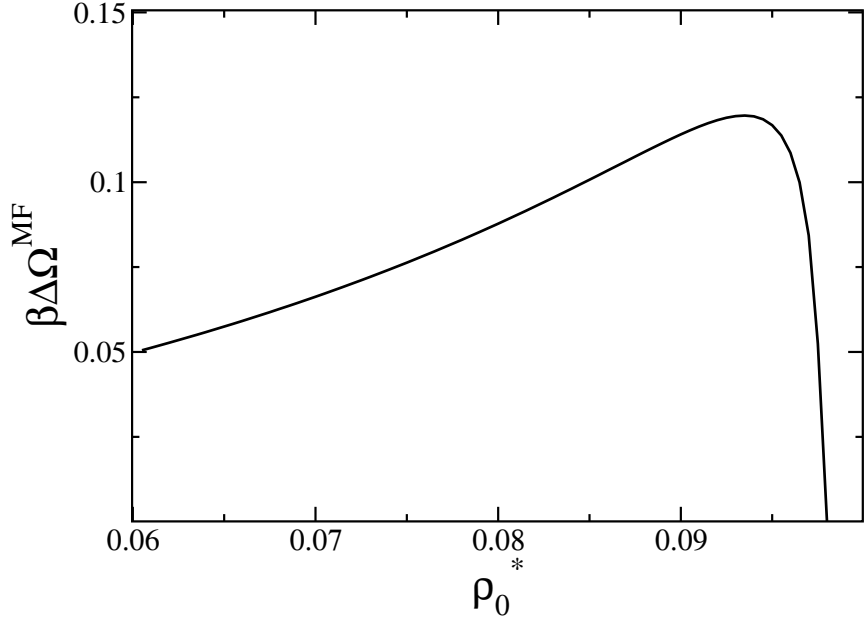}
\caption{The MF grand potential 
$\beta\Delta\Omega^{MF}[\Phi({\bf x}),\rho^*({\bf x})-\rho^*_0]$ (see
Eq.(\ref{delom})) along the approximate transition line
(\ref{trans_MF_line}), where ${\cal H}_{WF}=0$. $\Phi({\bf x})$ and
$\rho^*({\bf x})$ are given in Eq. (\ref{phii}) with (\ref{phi1a}),
and in Eq. (\ref{rho_space}) respectively. }
\end{figure}

\subsection{Effects of fluctuations on phase transitions}

In this subsection we include the fluctuation contribution to $\Omega$
in Eq.(\ref{OmHarval}).

We consider two cases, the $\phi^4$ theory (${\cal A}_6\equiv 0$ in
the above equations), as in the Brazovskii work \cite{brazovskii:75:0},
and the $\phi^6$ theory.
\subsubsection{Results of the $\phi^4$ theory}

The $\phi^4$ theory is stable for $\rho^*_0>\rho^*_{tc}$, and for such
densities the term $O(\phi^6)$ can be neglected. In this case
analytical solutions for $r_0$ and $r$ of Eqs.(\ref{C_r}) and (\ref{min_cond}) can be obtained.
 Physical solution for $r$ 
 corresponds
to the lowest value of $\Phi_1$. The transition lines
between the uniform and the two ordered phases are shown in Fig.6.  In
the one-shell approximation the lamellar order turns out to be more
stable than the P phase. We also considered the much more tedious
two-shell approximation for a few points. We found $\Phi_2$
significantly smaller than $\Phi_1$, and the phase-transition lines
shifted to somewhat lower temperatures compared to the one-shell
approximation. The rest of our results is obtained in the much simpler
one-shell approximation.
\begin{figure}
\includegraphics[scale=1.25]{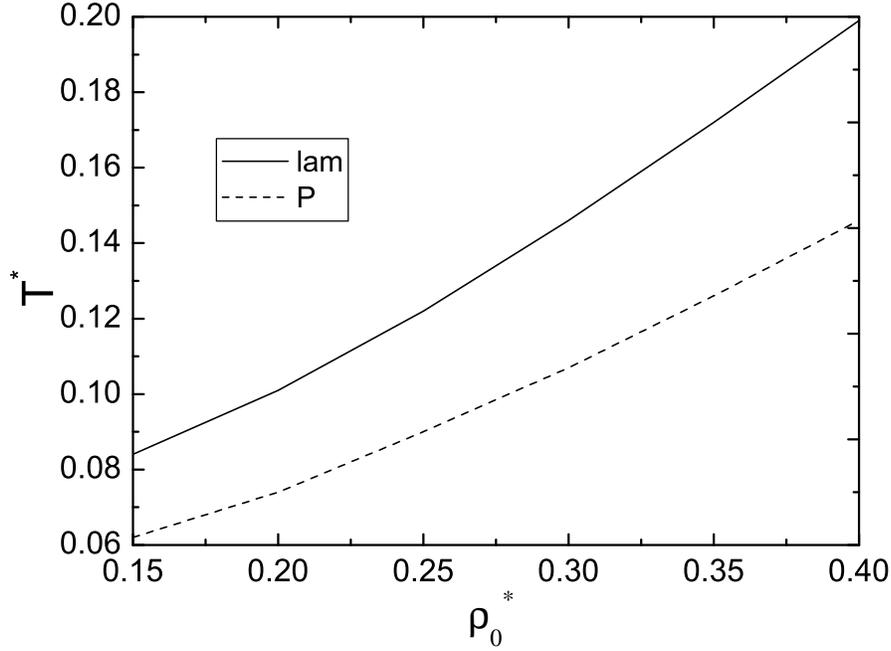}
\caption{The fluctuation-induced first-order transition lines between
the fused salt and the ordered phases in the $\phi^4$ theory in the
$(\rho_0^*,T^*)$ phase diagram. The transition to the P phase is
 metastable.}
\end{figure}

 Recall that our results rely on the approximate Eq.(\ref{calG1}),
which is valid provided that the condition $r\ll
v_2k^2_b/T^*(\rho_0^*)$ is satisfied. We verified that along the
coexistence lines $T^*(\rho_0^*)$ (Fig.6), the $r$ and $r_0$ are one
and two orders of magnitude smaller than $
v_2k^2_b/T^*(\rho_0^*)\approx 6/T^*(\rho_0^*)$, respectively. Thus,
close to the phase coexistence the approximation (\ref{calG1}) used
in
 our calculations is valid. However, the condition ${\cal
A}_4\sqrt{\beta^*v_2}k_b\ll r$, under which the  disregarded diagrams
(including Fig.2b) can be neglected~\cite{brazovskii:75:0}
is satisfied only for high densities. Therefore the accuracy of our results
increases with increasing density. 

\begin{figure}
\includegraphics[scale =1.2]{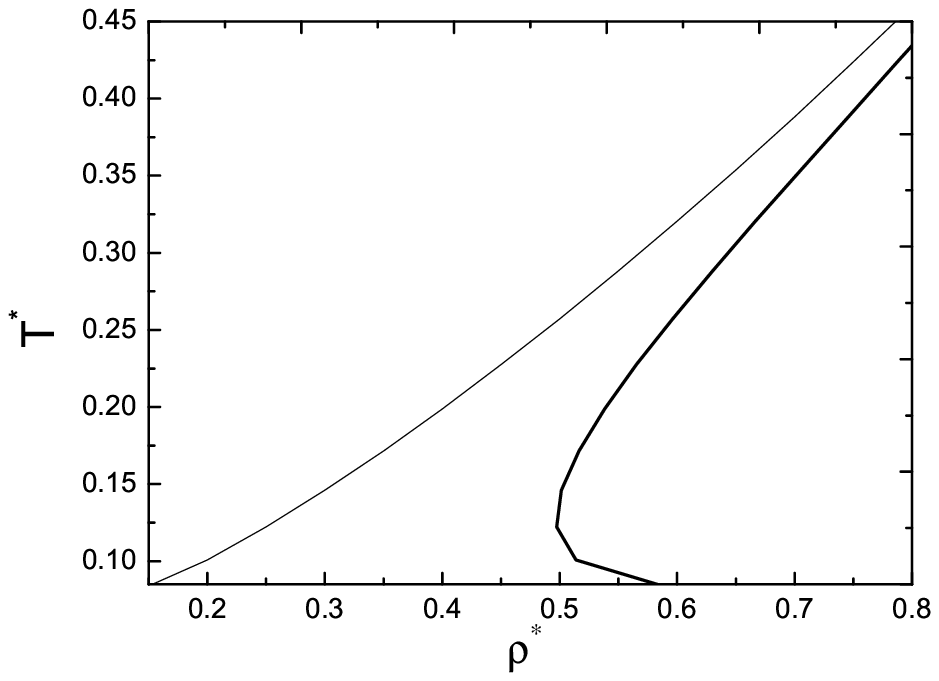}
\caption{Temperature at the liquid- lamellar phase
transition in the $\phi^4$ theory as a function of the average density
in the liquid
 phase at different levels of approximation in
Eq.(\ref{etav}). Thin solid line corresponds to the zeroth-order term
in Eq.(\ref{etav}), i.e. $\rho^*$ is approximated by $\rho^*_0$. Thick
solid line is obtained by including the leading-order contribution to
the fluctuation-induced density shift, i.e. $\rho^*$ is approximated
by $\rho^*_1$ (Eq.(\ref{der})). Explicit expression for the average
density in the liquid is given Eq.(\ref{rsh}). }
\end{figure}
In Fig.7 the temperature at the transition to the lamellar phase is
shown as a function of the most probable density (MF result) and as a
function of the average density, given by the approximate expressions
(\ref{der}) and (\ref{rsh}).
 The average density is a
 nonmonotonic
function of $T^*$ along the phase coexistence (Fig.7) for
$T^*<0.15$. As already discussed in sec.3a, for the corresponding
range of $\rho^*_0$ the approximate functional (\ref{phi6t}) is
oversimplified. Moreover, the neglected diagrams (Fig.2b) may yield a
relevant contribution to the grand potential for low densities.

Let us
compare the temperatures at the continuous transition in MF
 and at
the first-order transition in our theory (Figs.4 and 7). In
particular, for $\rho^*=0.8$ we find $T^*\approx 1.29$ and
$T^*\approx 0.43$ in the first and in the second case
respectively. For $T^*\approx 0.43$, on the other hand, we find in MF
the transition-density $\rho^*_0\approx 0.28$, a much lower value
than in our theory.  As we see, in this approximation the
fluctuation-induced shift of the liquid-phase boundary is
substantial. However, for $\rho^*=0.8$ the temperature at the
transition between liquid and the CsCl crystal obtained in
simulations ~\cite{vega:03:0} is $T^*\approx 0.1$.  Before we
identify the charge-ordered phase, we need to find out how the
transition temperature and density change when better approximations
 for ${\cal
H}_{WF}$, for the function $\Phi({\bf x})$ and for the average
density are made within our theory.  In the next section we study the
role of the $\phi^6$ term in Eq.(\ref{phi6t}).
 
\subsubsection{Results of the $\phi^6$ theory}

In this subsection we determine the effect of the $\phi^6$ term on the
phase behavior. Analytical solutions for the phase
transitions can be obtained by using the original Brazovskii method
\cite{brazovskii:75:0,fredrickson:87:0}, if in equations determining the
phase transition the terms of the highest order in $\Phi_1$ are
neglected. This is justified when $\Phi_1\ll 1$.  In Appendix D we
explain the key steps of the calculations. The full equations in the
$\phi^6$-theory can only be solved numerically.  Results for the
transition lines between the uniform and the two ordered phases are
shown in Fig.8, where analytical results of the approximate theory
and
 numerical results of the full $\phi^6$-theory are shown as lines
and
 as symbols respectively. The liquid-phase boundary,
$T^*(\rho^*_0)$, is shown in Fig.9 as a function of the average
density at different levels of approximation in Eq.(\ref{etav}). The
thick solid line is obtained from Eq.(\ref{etav}) with the two
leading-order terms included, i.e. in an approximation consistent with
Eq.(\ref{phi6t}). We verified that Eq.(\ref{calG1}) used in
 our
calculations is also valid in the $\phi^6$-theory.
\begin{figure}
\includegraphics[scale=1.25]{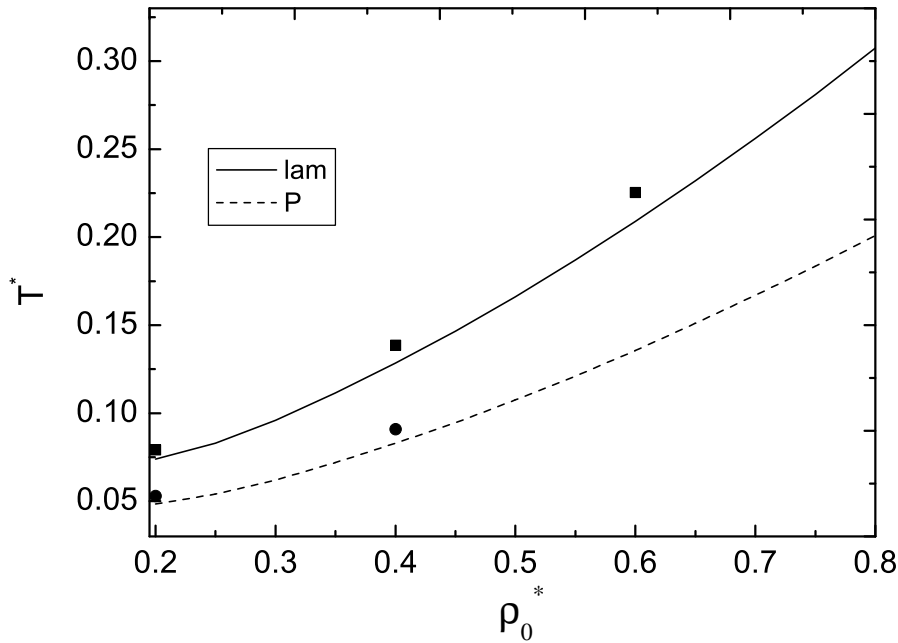}
\caption{The fluctuation-induced first-order transition lines between
the liquid- and the ordered phases in the $\phi^6$ theory in the
$(\rho_0^*,T^*)$ phase diagram. Lines are the analytical solutions of
the approximate equations (Appendix D) and symbols are the numerical
solutions of the full equations described in the text. The transition
to the P phase is metastable.}
\end{figure}

\begin{figure}
\includegraphics[scale =1.25]{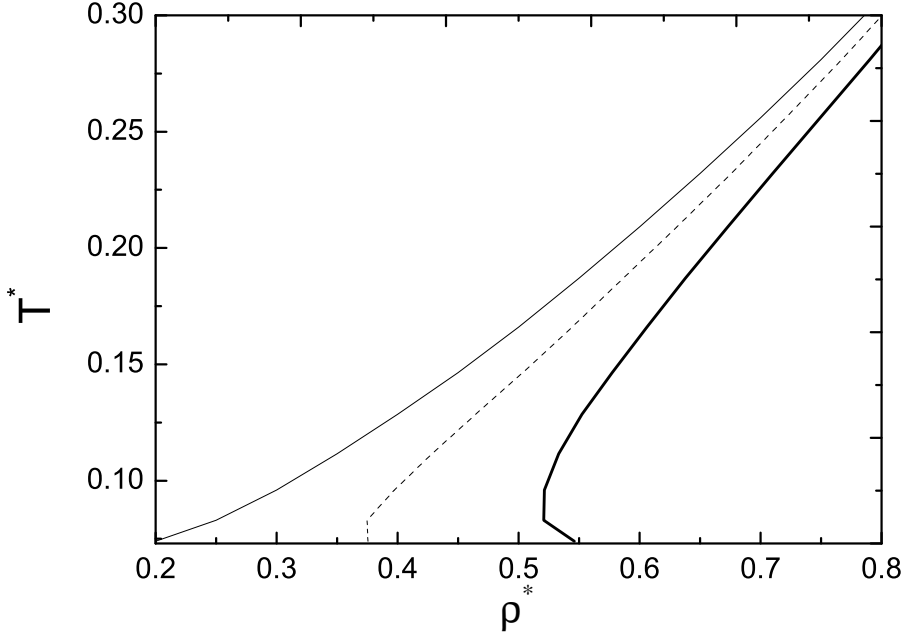}
\caption{Temperature at the  first-order
 phase
 transition between liquid and the charge-ordered phase in the
 $\phi^6$ theory, as a function of the average density (\ref{etav}) at
 different levels of approximation. Thin solid line corresponds to
 $\rho^*$ approximated by the MF result, $\rho^*_0$. Along the dashed
 line $\rho^*$ is given by the space-averaged leading-order
 fluctuation-contribution to the average density $\rho^*_1$ (
 Eq.(\ref{der})). Thick solid line corresponds to $ \rho^*_2 $ (
 Eq.(\ref{der2})), where the next-to-leading orderd term in
 (\ref{etav}) is taken into account.  }
\end{figure}

 By comparing Figs.6 and 8 we see that in the $\phi^6$-theory both
 transition lines are significantly shifted to lower temperatures
 compared to the $\phi^4$-theory. Fig.9 shows that in the consistent
 approximation for the average density, higher density at the phase
 coexistence is obtained. The results of the $\phi^6$ theory
 are thus
 closer to the simulation results for the liquid-CsCl
 transition, but
 the transition temperatures are still too high. For
 example, for
 $\rho^*=0.8$ we have $T^*\approx 0.43$ and $T^*\approx
 0.28$ in the
 $\phi^4$ and in the $\phi^6$ theory respectively, whereas
 for the
 liquid-CsCl transition $T^*\approx 0.1$ for the same density
 \cite{vega:03:0}.  Note that the MF results (see the discussion at
 the end of sec.4a) suggest that the truncated
 functional
 (\ref{phi6t}) leads to overestimated transition
 temperatures
 compared to the original functional
 (\ref{delom}).  We can expect
 that by addition of the term
 $\sim\phi^8$ in Eq.(\ref{phi6t}) and
 the third-order term in Eq.(\ref{etav}), lower temperatures and
 higher densities at the transition should be obtained. Better forms
 of the charge-density profile $\Phi({\bf
 x})$ should lead to lower
 transition-temperatures as well, as
 indicated by the (partial)
 results that we obtained in the two-shell
 approximation.  Thus,
 systematic improvement
 of the approximations that we make in
 explicit calculations, should
 lead to systematically decreasing
 transition temperature and
 increasing density, and our results
 should systematically approach
 the simulation results for the liquid
 -- ionic-crystal transition
 \cite{vega:03:0}.

For the two quite different charge-ordered phases the transitions to
the disordered phase are not far from each other (Fig.8). It is thus
plausible that the transitions to the other ordered structures,
including the stable one, are located in the same part of the phase
diagram.

\begin{figure}
\includegraphics[scale =1]{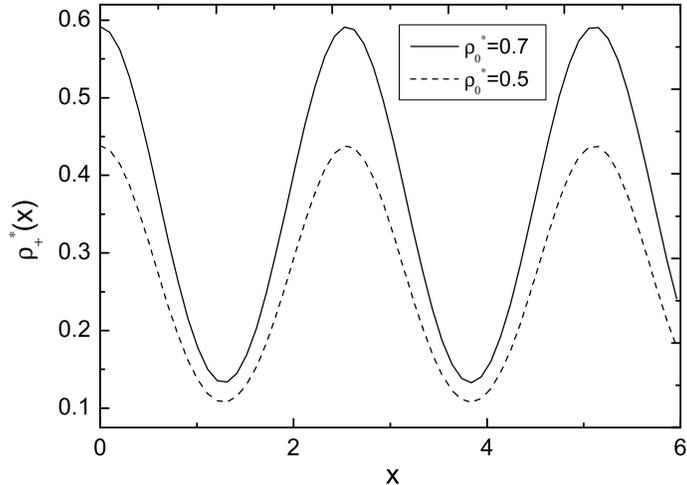}
\caption{The density profiles of cations,
$\rho^*_+(x)$,
 in the charge-ordered phase at the coexistence with
the liquid phase for two different
densities in the
 $\phi^6$ theory. Dashed line corresponds to $\rho_0^*=0.5$ ($\rho^*\approx
0.6$) and the solid line corresponds to $\rho^*_0=0.7$ ($\rho^*\approx
0.75$). $x$ is in $\sigma$-units and $\rho^*_+$ is
 dimensionless.}
\end{figure}

 Let us focus on the density difference between the coexisitng
phases. In the high-temperature part of the phase coexistence the
density difference $\Delta \rho^*$ between the coexisting phases is
rather small (see Ref.\cite{vega:03:0} and Fig.1).  By using
Eqs.(\ref{sp_av}) and (\ref{der2}) we find a vanishingly small
density
 difference between the coexisting liquid- and lamellar
phases. This
 result probably follows from the poor, one-shell ansatz
for the
 charge-density profile. The density profile of cations,
$\rho^*_+(x)=(\rho^*_2(x)+\Phi(x))/2$, where $\rho^*_2(x)$ is given in
Eq.(\ref{der2}), is shown in Fig.10 for two densities at the
coexistence with the liquid phase.  Beyond the effectively one-loop
approximation we expect some (probably weak) dependence of the unit cell of the
crystal on density, but its determination requires further studies. The shape of the
 density profile
in the charge-ordered phase resembles an average density
 profile in a
crystal (but only in one direction).
Of course in the one-shell approximation the
 crystalline structure
cannot be reproduced accurately. Our results for the
 liquid phase are
more accurate, because they do not depend on the form
 of $\Phi({\bf
x})$, which is the weakest point of our explicit results.
\section{summary}

In this work we considered the fluctuation contribution to the
grand-thermodynamic potential in Eq.(\ref{omOm}) within the field-theoretic
description of the RPM. Our main purpose was a determination of an
order, location and nature of the transition between the
charge-disordered and charge-ordered phases. We obtained an
approximate expression (Eqs.(\ref{diff}), (\ref{C_r}) and
(\ref{calG1})) for the grand-potential functional of the
charge-density profile $\Phi({\bf x})$ in the ordered phase. Our
approximation is based on the self-consistent, effectively one-loop
Hartree approximation, applied to the $\phi^6$ theory that was
derived for the RPM in Ref.{\cite{ciach:04:1}.

We found that in the continuum-space RPM the $\lambda$-line of
continuous transitions disappears when the charge-density fluctuations
are included. Instead, a first-order transition to a charge-ordered
phase appears. The range of temperatures and densities at the
first-order transition to the charge-ordered phase is  similar to
the range of temperatures and densities at the liquid - ionic-crystal
transition found in simulations
\cite{vega:03:0}. In the charge-ordered phase the average
charge-density exhibits oscillations with a period $\sim 2.5\sigma$
(beyond the one-loop approximation the
 period may be slightly
different),
 and with an amplitude $\sim 0.5$. Thus, our results
strongly indicate that
 the fluctuation-induced first-order
order-disorder transition
 should be identified with ionic
crystallization.
 We can conclude that the $\lambda$-line found in
different theories of the MF
 type is in fact a MF indication of ionic
crystallization. 
 
 The explicit results for the first-order phase
transition were
 obtained analytically within the simplest, one-shell
approximation for
 $\Phi({\bf x})$ (see (\ref{phii})), for two ordered
structures:
 one-dimensional, lamellar phase, and three-dimansional, P
phase shown
 in Fig.3.  We do not expect that the structure of the
charge-ordered
 phase can be correctly reproduced on this simple level
of
 approximation. Our approximate result for the crystalline
structure is only a
 very crude approximation.
 The transition lines
for the two different structures,
 however, are located close to each
other on the $(\rho^*,T^*)$ phase
 diagram. It is thus very plausible
that the transition line to the
 stable phase should be located near
the two transition lines, and
 conclusions concerning the approximate
location of the order-disorder
 transition are justified. 

In the effectively one-loop approximation we find the same lattice
constant of the CsCl crystal $2\pi/k_b$ for a range of
densities. Beyond the one-loop approximation we expect a weak
dependence of the lattice constant on density. This result and Fig.10
suggest that the number of defects (vacancies) in the crystal
coexisitng with the liquid (fused salt) increases with decreasing
$T^*$. The crystal melts either when at relatively low $T^*$ many
defects are present (low $\rho^*$), or when there are only few
defects, but $T^*$ is high.

Our results show a reasonable agreement with simulations
even for the very crude approximation for $\Phi({\bf x})$, and we
verified that by addition of further terms to Eqs.(\ref{phi6t}),
(\ref{der}) and
 (\ref{phii}) a better agreement should be obtained.
The
 accuracy of the results can be significantly improved within the
approach developed in this work by choosing a better ansatz for the
form of $\Phi({\bf x})$.  Note that the fluctuation
 contribution to
$\beta\Omega$ depends only on global characteristics
 of $\Phi({\bf
x})$, i.e. on integrals of $\Phi({\bf x})^{2n}$, where
 $n\le 3$ in
the case of the $\phi^6$ theory. Only the MF contribution
 depends on
a detailed shape of $\Phi({\bf x})$ through the term
 $\int_{\bf
x}\int_{\bf x'}
\Phi({\bf x})C^0_{\phi\phi}({\bf x}-{\bf x}')\Phi({\bf x}')=\int_{\bf k}
\tilde\Phi({\bf k})\tilde C^0_{\phi\phi}({\bf k})\tilde\Phi({\bf k}') $.
 Numerical determination of the structure of the charge-ordered phase
 is thus possible within our apporach, but it goes beyond the scope
 of the present work. We conclude that the field-theoretic approach
 developed in this work is suitable for a description of ionic
 crystallization on a semiquantitative level. 
\begin{acknowledgments}
 This work was supported by the KBN through a research project 1 P02B
 033 26.
\end{acknowledgments}
\section{Appendices}
\subsection{ Coefficients ${\cal A}_4,{\cal A}_6$ and $a_2$ 
in the WF approximation}
The coupling constants in the WF approximation are given in terms of
the coefficients $\gamma_{2m,n}$ \cite{ciach:04:1},
\begin{equation}
\label{hypA4}
{\cal A}_4=\gamma_{4,0}-3\frac{(-\gamma_{2,1})^2}{\gamma_{0,2}} ,
\end{equation}
\begin{equation}
\label{hypA6}
{\cal A}_6=\gamma_{6,0}-15\frac{(-\gamma_{2,1})(-\gamma_{4,1})}
{\gamma_{0,2}}
-15\frac{(-\gamma_{2,1})^3(-\gamma_{0,3})}{\gamma_{0,2}^3}-
45\frac{(-\gamma_{2,2})(-\gamma_{2,1})^2}{\gamma_{0,2}^2} ,
\end{equation}
 and in the CS
approximation they assume the explicit forms
\begin{equation}
\label{A4}
{\cal A}_4=-\frac{1-20s+10s^{2}-4s^{3}+
s^{4}}{{\rho_{0}^{*}}^{3}(1+4s+4s^{2}-4s^{3}+s^{4})}
\end{equation}

and

\begin{eqnarray}
\label{A6}
{\cal A}_6&=&\frac{3W(s)}{{\rho^{*}_{0}}^{5}(1+4s+4s^{2}-4s^{3}+s^{4})^{5} },
\end{eqnarray}
where 
\begin{eqnarray*}
&&W(s)=3-84s+360s^{2}+2644s^{3}+1701s^{4}-8736s^{5}
\\ \nonumber
&&+11240s^{6}-8304s^{7}+3861s^{8}-1164s^{9}+240s^{10}-36s^{11}+3s^{12}
\end{eqnarray*}
The coefficient $a_2$ in Eqs.(\ref{etav}) and (\ref{der2}) is
\begin{equation}
\label{a2}
\frac{a_2}{2}=-\frac{\Gamma^0_{4,1}}{4!\gamma_{0,2}}.
\end{equation}
where
\begin{equation}
\label{a2ga}
\Gamma^0_{4,1}=\gamma_{4,1}-\frac{6(-\gamma_{2,2})(-\gamma_{2,1})}
{\gamma_{0,2}}
-\frac{3(-\gamma_{3,0})(-\gamma_{2,1})^2}{\gamma_{0,2}^2}.
\end{equation}
 In the liquid
phase the explicit form of the average density (\ref{der}) is
\begin{equation}
\label{rsh}
\rho^{*}_1=\rho_{0}^{*}+
\frac{a(1-s)^4}{\rho_{0}^{*}(1+4s+4s^2-4s^3+s^4)}
\sqrt{\frac{T^{*}}{r_{0}}},
\end{equation}
 where Eq.(\ref{calG1}), Eq.(\ref{C_r}) with $\Phi=0$, and the CS
 reference system have been used.
 The explicit form of $\rho^*_2$ can
 be obtained in the same way with the help of Eqs.~(\ref{a2}),
 (\ref{a2ga}).
\subsection{ Explicit forms of the functions $C_{\phi\phi}^{fluc}$ and $C_i$ 
in the functional (\ref{Omfluc})}
\begin{equation}
\label{Cfluc}
C_{\phi\phi}^{fluc}({\bf x}-{\bf x}')=
C_{\phi\phi}^{0}({\bf x}-{\bf x}')+\Big(
\frac{{\cal A}_4}{ 2!}\Phi^2({\bf x})+
\frac{{\cal A}_6}{4!}\Phi^4({\bf x})\Big)\delta({\bf x}-{\bf x}'),
\end{equation}
\begin{equation}
\label{H}
C_1({\bf x})=\int_{\bf y}\Phi({\bf y})C_{\phi\phi}^{0}({\bf y}-{\bf x})
+\frac{{\cal A}_4}{3!}\Phi^3({\bf x}) +
\frac{{\cal A}_6}{5!}\Phi^5({\bf x}),
\end{equation}
\begin{equation}
\label{c3}
C_3({\bf x})={\cal A}_4\Phi({\bf x})+
\frac{{\cal A}_6}{6}\Phi({\bf x})^3,
\end{equation}
\begin{equation}
\label{c4}
C_4({\bf x})={\cal A}_4+
\frac{{\cal A}_6}{2}\Phi({\bf x})^2,
\end{equation}
and 
\begin{equation}
\label{c5}
C_5({\bf x})=\Phi({\bf x}).
\end{equation}
\subsection{ Explicit expressions for the phase transitions in 
the MF approximation}
The line of the first-order transition and the amplitude of the
charge-density wave are given by
\begin{eqnarray}
\label{trans_MF_line}
T^*=\frac{8{\cal A}_6\kappa_o\tilde V(k_b)\rho^*_0}
{5({\cal A}_4s_o)^2\rho^*_0-8{\cal A}_6\kappa_o}.
\end{eqnarray}
and
\begin{equation}
\label{phi1a}
\Phi_1^2=-\frac{{\cal A}_4s_o}{{\cal A}_6\kappa_o}
\end{equation}
respectively. 
The space-averaged density shift has the form
\begin{eqnarray}
\label{delro}
\Delta\rho^*=\frac{15\gamma_{2,1}{\cal A}_4s_o^{lam}}
{2\gamma_{0,2}{\cal A}_6\kappa_o^{lam}}.
\end{eqnarray}

\subsection{Explicit expression for the grand-potential difference}
After a substitution of Eqs.(27) and (28) into Eq.(22), the Brazovskii's 
equation of state [22],
\[
h=\frac{\delta \Omega}{\delta\tilde \Phi({{\bf k}_{b}})},
\]
can be written as
\begin{eqnarray}
\label{fin}
h&=&
\Big(\tilde C^0_{\phi\phi}({\bf k}_{b})+
\frac{{\cal A}_4{\cal G}}{2}+
\frac{{\cal A}_6{\cal G}^2}{8}\Big)\tilde \Phi(-{\bf k}_{b})\\ \nonumber
&&+\Big(\frac{{\cal A}_4}{3!}+\frac{{\cal A}_6}{12}{\cal G}\Big)
\int_{\bf k'}\int_{\bf k''}\int_{\bf k'''}
\delta({\bf k}_{b}+{\bf k'}+{\bf k''}+{\bf k'''})
\tilde \Phi({\bf k'})\tilde \Phi({\bf k''})\tilde \Phi({\bf k'''})\\ \nonumber
&&+\frac{{\cal A}_6}{5!}
\int_{\bf k'}\int_{\bf k''}\int_{\bf k'''}\int_{\bf k''''}\int_{\bf k'''''}
\delta({\bf k}_{b}+{\bf k'}+{\bf k''}+{\bf k'''}+{\bf k''''}+{\bf k''''})\prod_{i}
\tilde\Phi({\bf k}^i),
\end{eqnarray}
where $\tilde C^0_{\phi\phi}({\bf k})$ and ${\cal G}$ are given
in Eqs. (26) and (43) respectively.
For $\tilde C_{\phi\phi}(k)$ (see Eq.(27)) we obtain Eq.(\ref{C_rH}) with 
(\ref{C_r}).

For $|\Phi|\ll 1$ we truncate the expansions in $\Phi$ in Eqs.(\ref{fin}) and
(\ref{C_r}) at the $O(\Phi^3)$ and $O(\Phi^2)$ terms
respectively. As a result, the correlation function is given in
Eq.(\ref{C_r}) with the term $O(\Phi^4)$ neglected.
 The truncated Eq.(\ref{fin}),  and
 Eq.(\ref{C_r}) give for $h$ the result
\begin{eqnarray}
\label{h}
h=r\tilde\Phi(-{\bf k}_b)+\Big({\cal A}_4+\frac{{\cal A}_6{\cal G}}{2}\Big)
\int_{{\bf k}_1}\int_{{\bf k}_2}
\tilde\Phi({\bf k}_1)\tilde\Phi({\bf k}_2)\Big(\frac{1}{3!}
\tilde\Phi({\bf k}_b-{\bf k}_1-{\bf k}_2)-
\frac{1}{2}\tilde\Phi({\bf k}_b)\Big).
\end{eqnarray}
  In the one-shell
approximation the explicit form of the equilibrium condition $h=0$ is 
\begin{eqnarray}
\label{h=0}
r+\Big({\cal A}_4+\frac{{\cal A}_6\alpha}{\sqrt r}\Big)b_n\Phi_1^2=0
\end{eqnarray}
where $\alpha=a\sqrt T^*$, with $a$ given in Eq.({\ref{a}), and
\begin{eqnarray}
\label{en}
b_n=\frac{\sqrt{2n_1}}{3!}\int_{{\bf k}_1}\int_{{\bf k}_2}
\tilde g_1({\bf k}_1)\tilde g_1({\bf k}_2)
\tilde g_1({\bf k}_b-{\bf k}_1-{\bf k}_2)- \frac{1}{2}.
\end{eqnarray}

For both the lamelar and P structures  $b_n=-1/(4n_1)$ .

 The difference between the thermodynamic potential in the ordered and the
 disordered phases can be obtained in the same way as in
 Ref.\cite{fredrickson:87:0}, and we find
\begin{eqnarray}
\label{DelGam}
\Delta\Omega=\frac{{\cal A}_4b_n}{4}\Phi_1^4+\Omega_r,
\end{eqnarray}
where

\begin{eqnarray}
\label{DelGamr}
\Omega_r=\int_0^{\Phi_{1}}d\varphi\varphi\Big(r+
\frac{{\cal A}_6 b_n\alpha}
{\sqrt r}\varphi^2\Big),
\end{eqnarray}
\begin{eqnarray}
\label{f^2}
\varphi^2=\frac{2r^{3/2}-2\sqrt{r}\beta^{*}\tau_0-
{\cal A}_4\alpha}{{\cal P}}-\frac{\alpha}{\sqrt r},
\end{eqnarray}
and
\begin{eqnarray}
\label{fdf}
\varphi d \varphi=\Big(\frac{\sqrt r}{{\cal P}}+\frac{\alpha}{2r^{3/2}}
+\frac{{\cal A}_6\alpha\varphi^2}{2r{\cal P}}\Big)dr,
\qquad
{\cal P}={\cal A}_4\sqrt r+{\cal A}_6\alpha.
\end{eqnarray}

 After inserting (\ref{f^2}) and (\ref{fdf}) into (\ref{DelGamr}) we
obtain an integral which can be calculated analytically (e.c. with the
help of
 Mathematica).  In order to obtain $r$ and $\Phi_1$ at the coexistence, we
solve
 Eq. (\ref{C_r}) supplemented with the condition (\ref{h=0}). As
a result, we arrive at the explicit expressions for $\Delta\Omega$ and
$r$, which were used in a determination of the phase diagram shown in
Fig.9, but are too cumbersome to be presented here.


\end{document}